\documentclass[aps,prl,twocolumn,floatfix]{revtex4}
\usepackage{graphicx}
\usepackage{graphics}
\usepackage{dcolumn}
\usepackage{bm}
\usepackage{amsmath}

\begin{document}

\title{Metastable $\pi$-junction between an s$_\pm$-wave and an s-wave
superconductor}
\author{E. Berg$^1$, N. H. Lindner$^{2,3}$, T. Pereg-Barnea$^{3}$}
\affiliation{$^1$ Physics Department, Harvard University, Cambridge, MA 02138, USA}
\affiliation{$^2$ Institute of Quantum Information, California Institute of Technology, Pasadena, CA 91125, USA}
\affiliation{$^3$ Department of Physics, California Institute of Technology, 1200 E.
California Blvd, MC149-33, Pasadena, CA 91125, USA}
\date{\today}

\begin{abstract}

We examine a contact between a superconductor whose order
parameter changes sign across the Brillioun zone, and an ordinary,
uniform-sign superconductor. Within a Ginzburg-Landau type model,
we find that if the the barrier between the two superconductors is
not too high, the frustration of the Josephson coupling between
different portions of the Fermi surface across the contact can
lead to surprising consequences. These include time-reversal
symmetry breaking at the interface and unusual energy-phase
relations with multiple local minima. We propose this mechanism as
a possible explanation for the half-integer flux quantum
transitions in composite niobium–-iron pnictide superconducting
loops, which were discovered in a recent
experiment\cite{Chen2010}.

\end{abstract}

\maketitle



\emph{Introduction--} Understanding the structure of the order
parameter of the iron-based pnictide
superconductors\cite{Kamihara2008} is the key to unveil their
pairing mechanism. A conventional, phonon-mediated mechanism is
usually associated with an order parameter of a uniform sign,
while sign changes in the order parameter are typical of
unconventional mechanisms, in which pairing is driven by purely
repulsive Coulomb interactions.

In the cuprates, the unambiguous identification of the d-wave
symmetry of the order parameter came primarily from
phase-sensitive experiments\cite{Wollman1993,Tsuei1994}. These
experiments exploit the fact that due to the d-wave symmetry,
specific geometries (such as a corner junction with an s-wave
superconductor, or a tri-junction between three d-wave
superconductors) are guaranteed to produce a $\pi$ phase shift in
the phase of the superconducting order parameter. In the iron
arsenides, similar experiments\cite{Hicks2008,Zhang2009} have
found no evidence for d-wave symmetry.

From the theory side, it has been proposed that the pnictides have
an extended s-wave (``s$_\pm$ wave'') order
parameter\cite{Mazin2008,Cvetkovic2009}, which can be chosen to be
real but changes sign between the electron and hole pockets, and
is invariant under the overall tetragonal symmetry (A$_{1g}$).
Such a pairing state has been found from solving both weakly
\cite{Kuroki2008,Chubukov2008,Graser2009,Wang2009} and strongly
\cite{Seo2008,Berg2010,Yang2010} interacting models. Although
there exist several experimental indications that this is indeed
the correct pairing state in certain pnictide
superconductors\cite{Christianson2008,Chi2009,Hanaguri2010,Teague2011},
more direct experimental evidence is highly desirable. Designing a
Josephson interferometry device which could detect the s$_\pm$
state poses a significant challenge, since symmetry alone does not
guarantee a $\pi$ phase shift in \emph{any} geometry. Several
ideas have been proposed to overcome this
difficulty\cite{Parker2009,Linder2009,Chen2009,Tsai2009,Chen2010b},
but none were realized to date.

Progress has been made recently, in the work of Chen \emph{et al.}\cite%
{Chen2010}. In this experiment, the flux through a composite
Niobium(Nb)-NdFeAsO$_{0.88}$F$_{0.12}$(FeAs) superconducting loop
was measured. By using an external electromagnetic pulse, is was
shown that both integer and half-integer flux jumps can be induced
in the loop, in units of the superconducting flux quantum
$\Phi_0=hc/2e$. While providing a strong indication for a sign
change in the order parameter, these results are surprising,
because neither a $\pi$-junction nor a $0$-junction between the Nb
and the FeAs superconductors would lead to the possibility of
half-integer flux quantum jumps.

Motivated by these experiments, we study a model of a junction
between a sign-changing and a conventional (s-wave)
superconductor. The model allows us to interpolate between the
tunneling (weak coupling) and the metallic contact (strong
coupling) regime. We find that above a certain critical coupling
strength, the frustration of the Josephson coupling across the
barrier can lead to unusual energy-phase relations in the
junction. If the phase stiffness of the superconductors is small
enough, the energy-phase relation has two minima which break
time-reversal symmetry; if the phase stiffness is large, the
global minimum of the energy is at a phase
difference of $\Delta \varphi=0$, but an additional meta-stable minimum at $%
\Delta \varphi=\pi$ appears. The latter situation can explain the Chen \emph{%
et al.} experiment, since it allows allows the junction to switch from $%
\Delta \varphi=0$ to $\pi$, causing a half-flux quantum jump in
the loop. The fact that the half-integer jumps appear only beyond
a certain value of the critical current in the loop, as well as
the relatively small probability of half-integer
jumps\cite{Chen2010}, can both be understood within our model.

\begin{figure}[t]
\includegraphics[width=\columnwidth]{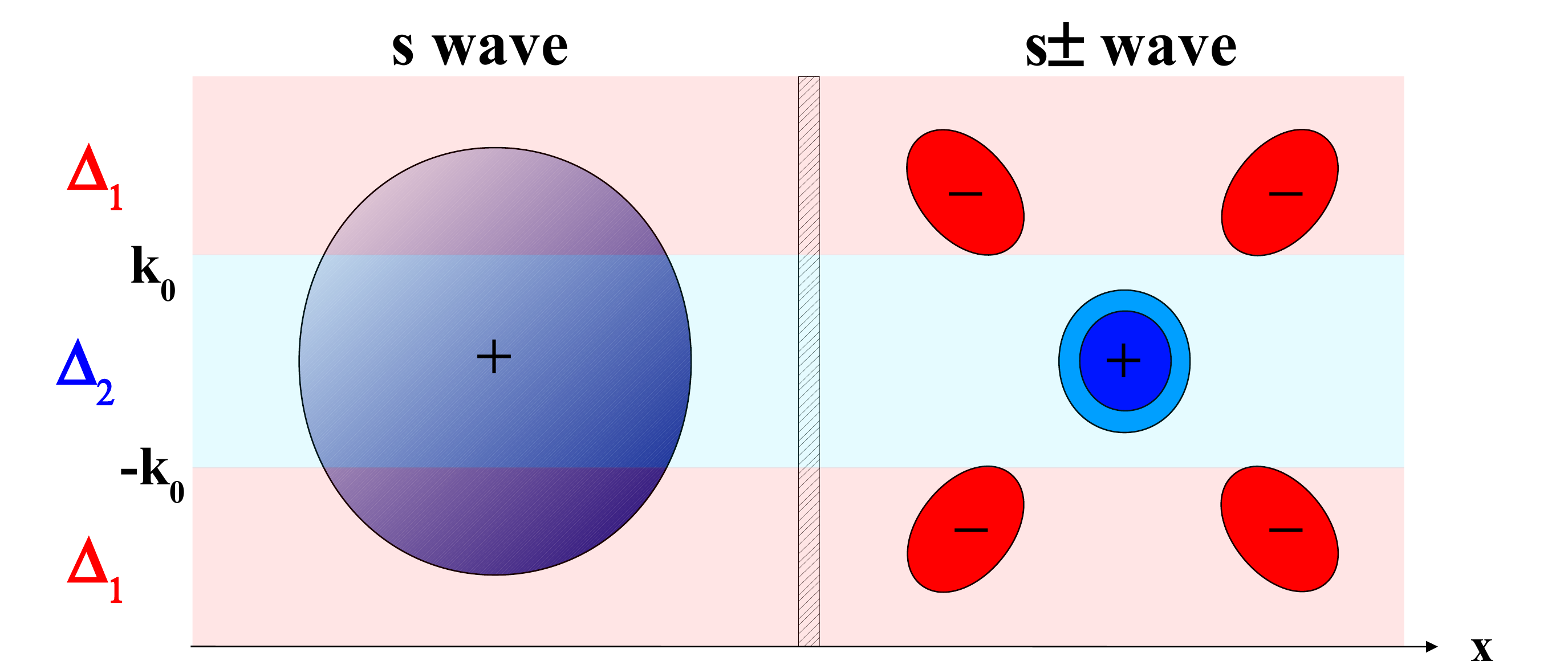}
\caption{(Color online.) Junction between an s-wave and an s$_\pm$
wave superconductor. We define two order parameters,
$\Delta_{i=1,2}$ (see Eq. \ref{eq:Delta}), which belong to
different regions in momentum space. In the s$_\pm$ side,
$\Delta_{1,2}$ have an opposite sign.} \label{fig:junction}
\end{figure}

\emph{The model--} We consider a planar Josephson junction between
an s-wave and a sign-changing s-wave, shown in Fig.
\ref{fig:junction}. Since we are interested in the crossover from
a tunneling barrier to a metallic contact, in which the order
parameters of both superconductors are modified significantly at
distances of the order of a few coherence lengths from the junction~\cite%
{DeGennes1964}, the problem needs to be treated self-consistently.
Rather than solving the full Bugoliubov-de Gennes equations, we
use an effective Ginzburg-Landau type free energy functional which
depends on the superconducting order parameter near the
junction~\cite{Sols1994}. Although the Ginzburg-Landau description
is strictly valid only close to the critical temperature
($T_\mathrm{c}$) of both superconductors, we expect it to capture
the qualitative behavior of the system even at lower temperatures.

In order to capture the multi-band nature of the system, we
introduce two superconducting order parameters, $\Delta _{i}$
where $i=1,2$. Microscopically, these can be viewed as belonging
to regions of different momenta parallel to the junction:
\begin{equation}
\Delta _{i}\left( x\right) =\frac{1}{A} \sum_{ k_{\parallel } \in
\Omega _{i}}V_{k_{\parallel },k_{\parallel }^{\prime
}}\left\langle \psi _{k_{\parallel }^{\prime }\uparrow }\psi
_{-k_{\parallel }^{\prime }\downarrow }\right\rangle ,
\label{eq:Delta}
\end{equation}%
where $\psi _{k_{\parallel }\sigma }^{\dagger }(x) $ is the
electron creation operator at position $x$, momentum $k_{\parallel
}$ parallel to the junction, and spin $\sigma$, $V_{k_{\parallel
},k_{\parallel }^{\prime }}$ is the pairing interaction in the
Cooper channel, $x$ is the coordinate perpendicular to the
junction, $A$ is the area of the junction, and the two momentum
regions $\Omega _{i}$ are defined by $\Omega _{1}=\left\{
k_{\parallel }\Big| k_{0}>\left\vert k_{\parallel }\right\vert
\right\} $ and $ \Omega _{2}=\left\{ k_{\parallel }\Big| k_{0}\leq
\left\vert k_{\parallel }\right\vert \right\} $, where $k_{0}$ is
an arbitrary momentum chosen such that in the s$_\pm$ side,
$\Delta _{1}>0$ and $\Delta _{2}<0$. There, $\Delta _{1}$ and
$\Delta _{2}$ can be thought of as the order parameters on
different bands. Note that such as a decomposition is possible
irrespective of the relative orientation of the two crystals, as
long as in the sign-changing s-wave side, the region $\Omega_1$
($\Omega_2$) is dominated by the positive (negative) part of the
order parameter.

We describe the junction using the following phenomenological
Ginzburg-Landau free energy:
\begin{equation}
F\left[ \Delta _{1},\Delta _{2}\right]
=F_{L}+F_{R}+F_{\mathrm{c}}\text{.} \label{eq:F}
\end{equation}
Here,
\begin{eqnarray}
F_{\nu }[\Delta_1,\Delta_2] &=&\int_{\nu} dx  \Bigg\{
\sum_{i=1,2}\Big[\frac{1}{2} \kappa _{i}^{\nu } \left\vert
\partial _{x}\Delta _{i}\right\vert ^{2}-\frac{1}{2} r_{i}^{\nu }\left\vert
\Delta _{i}\right\vert ^{2}  \notag \\
&&+\frac{1}{4}u_{i}^{\nu }\left\vert \Delta _{i}\right\vert ^{4}+{\frac{1}{4}%
}\alpha _{i}^{\nu }\left( \Delta _{i}^{\ast }\partial _{x}\Delta _{i}-\Delta
_{i}\partial _{x}\Delta _{i}^{\ast }\right) ^{2}\Big]  \notag \\
&&-v^{\nu }(\Delta _{1}^{\ast }\Delta _{2}+c.c.)\Bigg\} \text{,}
\label{eq:Fnu}
\end{eqnarray}
where $F_{\nu =L,R}$ are the free energies of the left (Nb) and
right (FeAs) sides, with $\int_{L}\equiv \int_{-\ell}^{0}$ and
$\int_{R}\equiv \int_{0}^{\ell}$ ($2\ell$ is the system length),
and
\begin{equation}
F_{\mathrm{c}}\left[ \Delta _{1},\Delta _{2}\right]
=\sum_{i=1,2}\left[ T_{i}\left\vert \Delta _{i}\left( 0^{+}\right)
-\Delta _{i}\left( 0^{-}\right) \right\vert ^{2}\right]
\label{eq:Fc}
\end{equation}%
describes the contact between the two superconductors. In some
cases, we will consider a ``ring'' geometry in which we add to
$F_\mathrm{c}$ another term, $\sum_{i=1,2}\left[ T_{i}\left\vert
\Delta _{i}\left( \ell \right) -\Delta _{i}\left( -\ell \right)
\right\vert ^{2}\right]$.

$\kappa^\nu _{i}$,$r^\nu_{i}$,$u^\nu_{i}$ in Eq. \ref{eq:Fnu} are
the standard Ginzburg-Landau parameters of band $i=1,2$ on the
left/right ($\nu=L,R$ respectively) of the
junction\cite{Comment-other-terms}. The $\alpha^\nu_{i}$ terms
represent an additional energy cost of creating supercurrents.
These terms turn out to be necessary to describe the transition
from a single-minimum to a double-minimum junction (see below).
Close to the critical temperature, the $\alpha$ term becomes
negligible compared to the other terms in Eq. \ref{eq:Fnu}, since
it is of higher order in the $\Delta_i$'s and their derivatives.
At lower temperatures, however, it can become important.

The parameter $v^\nu$ describes the inter-band coupling, and
encodes the tendency towards s or s$_{\pm }$ pairing: positive
(negative) $v^{L,R}$ corresponds to an s-- (s$_\pm$--) wave
superconductor, respectively.

$T_i$ (Eq. \ref{eq:Fc}) represent the strengths of the couplings
of the two order parameters across the barrier. These parameters
allow us to interpolate between the tunneling regime ($T_i
\rightarrow 0$) to the metallic contact regime ($T_i\rightarrow
\infty$). Note that in the latter regime, $\Delta_{1,2}$ become
continuous at the junction. This is the boundary condition of a
metallic contact between two superconductors, assuming for
simplicity that the density of states at the Fermi energy times
the pairing interaction is continuous at the
contact~\cite{DeGennes1964}. Our results do not depend
qualitatively on this assumption. 

\emph{Results--} We minimize the free energy (Eq. \ref{eq:F})
numerically. The minimization is done on a discrete lattice, with
a small enough lattice spacing such that the results are
independent of its size. In order to map the energy-phase relation
of the s-s$_\pm$ junction, we use the following boundary
conditions at $x=\pm \ell$: $\arg{\Delta_1(-\ell)=0}$,
$\arg{\Delta_1(\ell)=\Delta \varphi}$, where $\varphi$ is varied
between $0$ and $\pi$. The results (in particular, the qualitative
behavior of the junction) do not depend on the choice of $\ell$.


\begin{figure}[t]
\includegraphics[width=1.15\columnwidth]{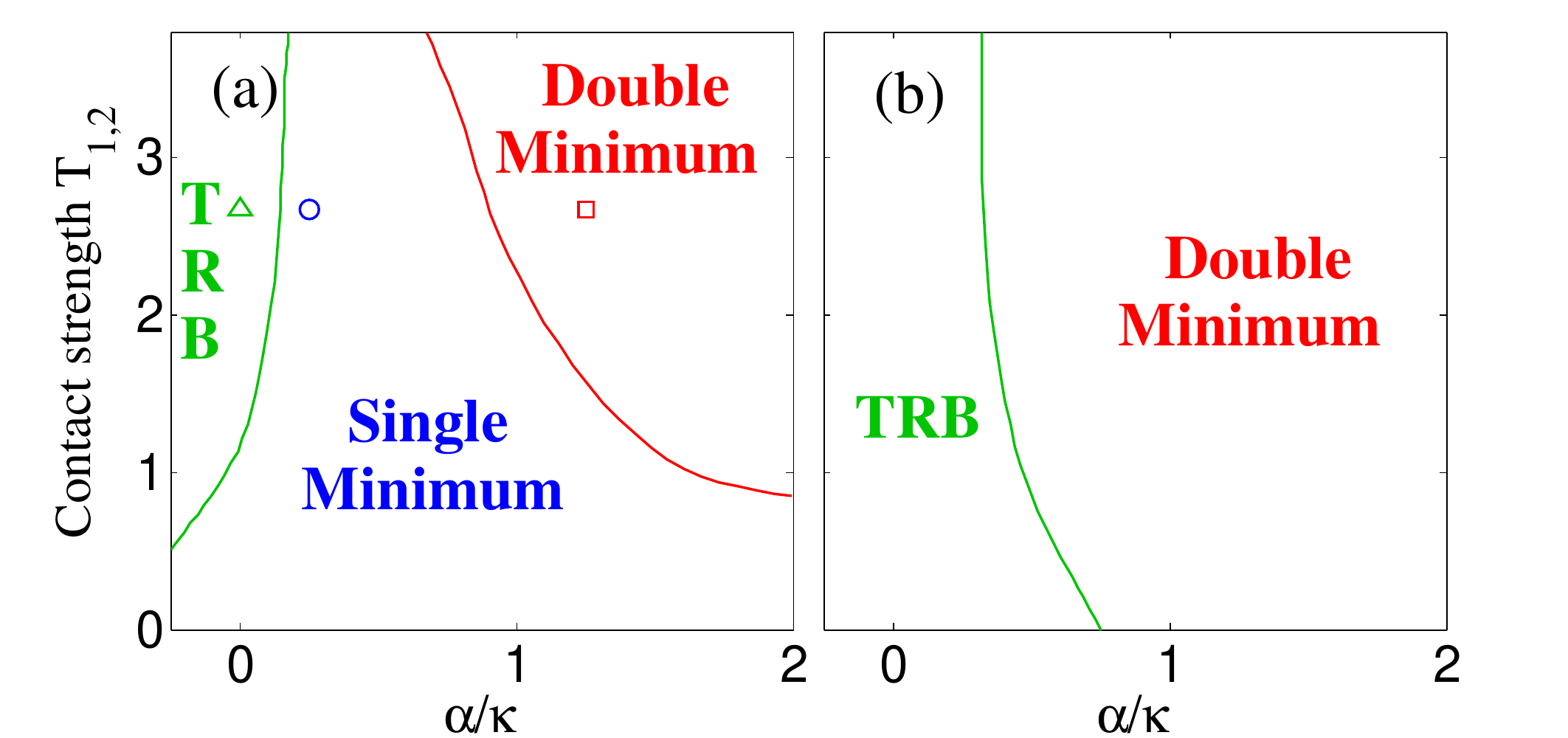}
\caption{(Color online.) (a) Phase diagram of an $s-s_\pm$
junction, using $r^\nu_i=1$, $\kappa^\nu_i=4$, $u^\nu_1=1$,
$u^\nu_2=2$, $v^L=-v^R=1$, $\alpha^\nu_i=\alpha$, $\ell=3$ and
$T_1=T_2$. The different states correspond to different
qualitative behaviors of the energy-phase relation in the
junction: a ``single minimum'' state in which the minima of the
free energy occur at a phase difference of $2n\pi$, a
``time-reversal breaking'' (TRB) phase in which the minima occur
at $2n\pi \pm \Delta \varphi_0$, and a ``double minimum'' phase in
which there are minima at $n\pi$. The triangle, the circle and the
square correspond to the three parameter sets which are used in
Fig. \ref{fig:FE}. (b) Same as (a), for a case where the
parameters are tuned such that $\Delta_{1,2}$ are equivalent (by
setting $u^\nu_1=u^\nu_2=1$). In this case, the single minimum
phase does not occur.} \label{fig:PD}
\end{figure}

Depending on the contact strength and on various material
parameters, we find three qualitatively different regimes:
1. A ``single minimum'' regime, which is realized for small $T_i$,
in which the free energy is minimal at $\Delta \varphi=2n \pi$
(where $n$ is an integer); 2. A ``time reversal breaking'' (TRB)
regime\cite{comment-TRB} at intermediate $T_i$ and small $\alpha$,
where the free energy exhibits degenerate minima at $\Delta
\varphi=2n \pi \pm \varphi_0$; 3. A ``double minimum'' regime, in
which there are global minima at $\Delta \varphi=2n \pi$ and local
minima at $\Delta \varphi=(2n+1) \pi$. This regime is realized for
sufficiently large $\alpha$ and $T_i$.

The phase diagram as a function of $\alpha$ and $T_i$, showing the
three phases described above, is shown in Fig. \ref{fig:PD}a. The
parameters used are listed in the figure caption.

In order to understand these results, we consider an artificial
situation in which the parameters of the free energy are tuned
such that the two order parameters $\Delta_{1,2}$ are exactly
equivalent. In this case, the free energy of the junction
$F(\Delta \varphi)$ (after minimization over $\Delta_{1,2}$) is
invariant under a shift of $\Delta \varphi$ by $\pi$. Together
with time-reversal symmetry, $F(\Delta \varphi)=F(-\Delta
\varphi)$, this dictates that there are two generic situations:
$F(\Delta \varphi)$ is minimal for either $\Delta \varphi=n\pi$ or
$\Delta \varphi=(n+1/2)\pi$. This can be understood as follows:
due to the sign change of the order parameter in the FeAs side,
the Josephson coupling is frustrated. In order to relieve this
frustration, the system can either twist the relative phase of
$\Delta_{1,2}$ close to the interface, or deform the amplitudes of
$\Delta_{1,2}$ such that $|\Delta_{1}|\ne |\Delta_{2}|$. The first
effect favors $\Delta \varphi=(n+1/2)\pi$, and the second favors
$\Delta \varphi=n\pi$. Which effect dominates depends on the
energy cost of a non-uniform phase relative to that of a
non-uniform amplitude of the order parameter. Increasing
$\alpha/\kappa$ increases the phase stiffness, and hence drives a
transition between the two regimes. The phase diagram for the case
of equivalent $\Delta_{1,2}$ is shown in Fig. \ref{fig:PD}b.

Upon making $\Delta_{1,2}$ inequivalent, the period of $F(\Delta
\varphi)$ becomes $2\pi$. However, the local minima of $F(\Delta
\varphi)$ remain stable over a finite region in parameter space.
Therefore, the two phases described above survive. Between them, a
third phase with minima at $\Delta \varphi=2n\pi$ appears.

\begin{figure}[t]
\includegraphics[width=0.82\columnwidth]{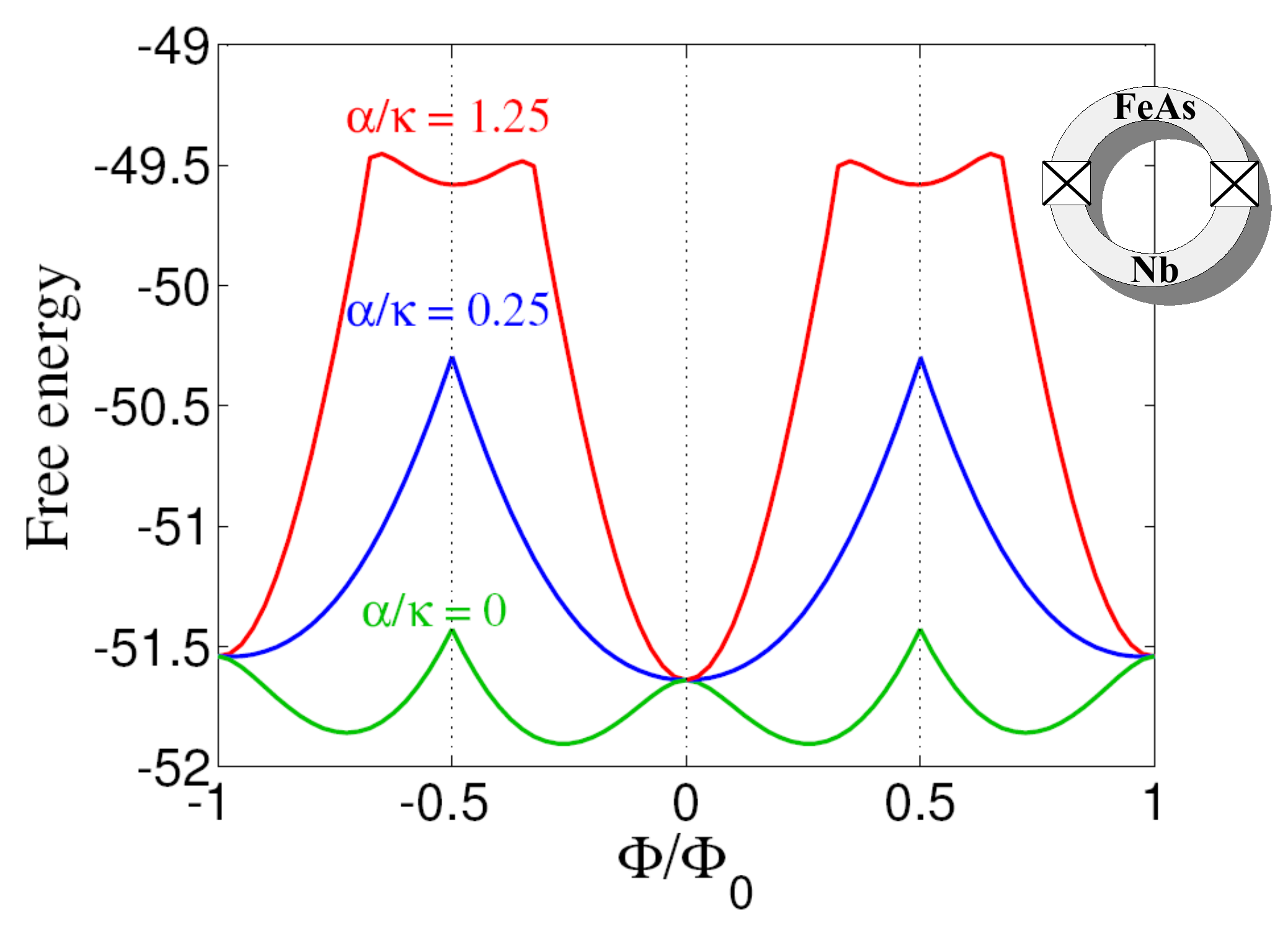}
\caption{(Color online.) Free energy as a function of flux in a
composite s--s$_\pm$ loop (inset) of length $2\ell=20$, for
$T_{1,2}=2.67$ and $\alpha/\kappa=0,0.25,1.25$. These correspond
to the ``time-reversal breaking'', ``single minimum'' and ``double
minimum'' states, respectively. The self-inductance of the loop
was chosen such that $\Phi_0^2/2L=0.1$.} \label{fig:FE}
\end{figure}

A composite s-s$_\pm$ superconducting loop (shown schematically in
Fig. \ref{fig:FE}) is also described by three qualitatively
different regimes, corresponding to the three phases of a single
junction. The free energy of the loop as a function of the flux
through the loop is given by\cite{Tsuei2000}
\begin{equation}
F_{\mathrm{loop}}=F(\Phi)+\frac{\Phi^2}{2L}\text{,}
\end{equation}
where $F(\Phi)$ is the free energy of Eq. \ref{eq:F} with
$\partial_x \rightarrow \partial_x - 2eA/h$ ($A=\Phi/2\ell$ is the
vector potential), supplemented with an additional contact term
(of the form of Eq. \ref{eq:Fc} with $x=\ell$), and $L$ is the
self-inductance of the loop. Fig. \ref{fig:FE} shows the free
energy of the loop as a function of $\Phi$,
with parameters $(\alpha,T)$ which place the system in one of the
three distinct regimes described above. For intermediate values of
$\alpha/\kappa$, the free energy of the loop has minima at
approximately $\Phi=n\Phi_0$ where $n$ is an integer (``single
minimum'' regime). For smaller $\alpha/\kappa$, each minimum
splits into two degenerate minima at $\Phi=n\Phi_0 \pm \Delta
\Phi$ where $\Delta
\Phi$ depends continuously on the various parameters of the
junction (TRB regime). Finally, for $\alpha/\kappa$ larger than a
critical value, minima appear both at $\Phi=n\Phi_0$ and
$\Phi=(n+1/2)\Phi_0$ (``double minimum'' regime).

\emph{Comparison to the Chen et al. experiment--} The experiment
of Chen \emph{et al.}\cite{Chen2010} can be interpreted in terms
of our model as follows. As the critical current in the composite
Nb--FeAs loop is increased, the Nb--FeAs junction undergoes a
transition from a single minimum to a double minimum phase. Upon
entry to the double minimum phase, both $n\Phi_0$ and
$(n+1/2)\Phi_0$ flux jumps are observed, and the probability for
$(n+1/2)\Phi_0$ jumps increases with increasing critical current
(which corresponds to increasing $T_{1,2}$ in Fig. \ref{fig:PD}).

Whenever the composite loop is excited electromagnetically, the
flux through the loop acquires a random ``kick'', and the system
can jump from one local minimum to another. The probability of
ending at a metastable local minimum, $\Phi=(n+1/2)\Phi_0$, is
smaller than the probability of ending in a global minimum,
$\Phi=n \Phi_0$. Therefore, if the system jumps from $n_1\Phi_0$
to $(n_2+1/2)\Phi_0$, the next jump is likely to be to $n_3\Phi_0$
(with integer $n_1$, $n_2$, $n_3$). This implies that the half
flux quantum jumps are correlated, and tend to appear in pairs.
Such a trend is clearly visible in the experimental results
\cite{Chen2010,Supplementary}. We conclude that the experimental
results of Ref. \cite{Chen2010} can be naturally explained by the
appearance of a metastable local minimum in the energy-phase
relation of the Nb-FeAs junction at $\Delta \varphi = (n+1/2)\pi$,
corresponding to the ``double minimum'' phase in our model.

\emph{Conclusions--} We have presented a model for a Josephson
junction between a simple s--wave and a sign-changing s$_\pm$ wave
superconductor. Due to the sign change in the s$_\pm$ side, the
Josephson coupling across the junction is partially frustrated. If
the barrier between the two superconductors is low enough,
corresponding to a metallic contact, the system can relieve this
frustration by either breaking time-reversal symmetry, or
developing additional local minima at a phase difference of
$(2n+1)\pi$. The half-flux quantum jumps in the experiment of Chen
\emph{et al.} can be explained by the second scenario. (Note that
a time-reversal breaking junction would correspond to fractional
flux jumps which are neither $n\Phi_0$ nor $(n+1/2)\Phi_0$, which
were not observed.)

It is important to note that this behavior is \emph{unique} to a
junction between an s--wave and a sign changing superconductor. In
a junction between two s--wave superconductors, only the single
minimum phase is realized, for \emph{any} strength of the coupling
$T_{1,2}$. To verify this, we considered a case in which
$v^L=v^R>0$, \emph{i.e.} both sides of the junction are s-wave
superconductors, and found only a single-minimum phase. In this
respect, the observation of half-flux quantum jumps is a strong
indication of a sign-changing order parameter. Together with the
lack of observation of spontaneous flux in a polycrystalline
sample\cite{Hicks2008}, which essentially rules out a d-wave order
parameter, we conclude that the s$_\pm$ is the most likely
candidate for the order parameter of F-doped NdFeAsO.

The presence local minima in the energy-phase relation of a
contact between a conventional and an iron-based superconductor
can be looked for in experiments. The AC Josephson effect should
reveal pronounced high harmonics of the AC current as a function
of the bias voltage. In a composite loop, of the same kind as the
one studied by Chen \emph{et al.}, half-flux quantum entries
should be observed as a function of an external field.

On the theoretical side, it will be interesting to examine a
strongly-coupled s--s$_\pm$ junction in a microscopic model.
Within our phenomenological model, we found that the fate of the
junction in the strong coupling limit is determined by the ratio
of the phase and amplitude stiffnesses. In a
Bardeen-Cooper-Schrieffer superconductor at low temperatures,
these stiffnesses both scale in the same way -- their magnitude is
proportional to the Fermi energy. Therefore, the relative
magnitude of the two stiffnesses is sensitive to microscopic
details. It will be particularly interesting to look for parameter
regimes in which the broken time-reversal junction can arise.

\emph{Acknowledgements--} The authors thank C. Hicks, S. Kivelson,
D. Podolsky, S. Raghu and D. Scalapino for useful discussions. We
are particularly grateful to C.-T. Chen for sharing her
unpublished data with us. EB was supported by the NSF under Grants
DMR-0705472 and DMR-0757145, NHL was supported by the Rothschild
Foundation and the Gordon and Betty Moore Foundation, and TPB was
supported by the Research Corporation Cottrell Scholars Award, and
DARPA. This research was supported in part by the NSF under Grant
PHY05-51164. EB and NHL thank the Aspen Center for Physics, where
part of this work was done.

\bibliography{mazleg}

\begin{thebibliography}{30}
\bibitem[{Com()}]{Comment-other-terms}
\bibinfo{note}{In Eq. 3, we have not included all the possible quartic terms
  which are allowed by symmetry. These terms are not expected to lead to any
  qualitatively new effects.}
\end{thebibliography}

\begin{widetext}

\section{Supplementary information: pairing of half-flux jump events}


\begin{figure*}[b]
\includegraphics[width=\textwidth]{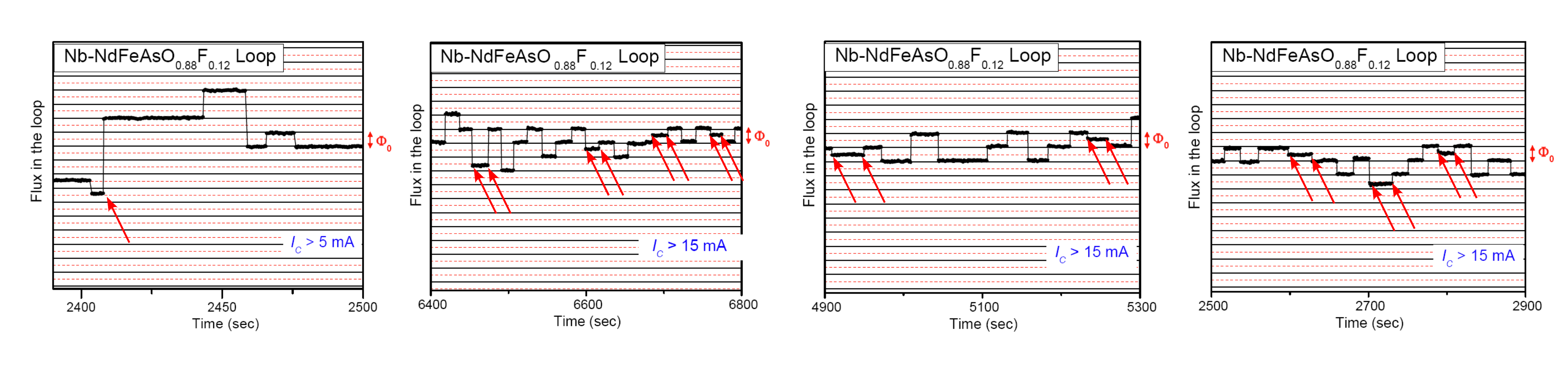}
\caption{Flux vs. time in a composite
Nb-NdFeAsO$_{0.88}$F$_{0.12}$ from Chen et. al.\cite{Chen2010}.
The abrupt jumps in the flux are caused by external
electromagnetic pulses. In each jump, the flux changes by either
$n\Phi_0$ or $(n+1/2)\Phi_0$, where $n$ is an integer and
$\Phi_0=hc/2e$. The arrows indicate the position of the
$(n+1/2)\Phi_0$ jumps.} \label{fig:flux_vs_time}
\end{figure*}

In Fig. \ref{fig:flux_vs_time}, we show scans of the flux vs. time
from the Chen et. al. experiment\cite{Chen2010}. The arrows
indicate the position of fractional $(n+1/2)\Phi_0$ flux jumps.
From the data, it is apparent that the $(n+1/2)\Phi_0$ jumps are
\emph{paired}, i.e., a half-integer flux jump tends to be followed
by another half-integer jump. This is consistent with the picture
of a local minimum in the energy vs. flux relation of the loop at
$(n+1/2)\Phi_0$. After each jump, the flux is less likely to get
trapped in a local minimum than in a global minimum at $n\Phi_0$.
Therefore, a jump from a global to a local minimum (which is
accompanied by a change in the flux by $(n+1/2)\Phi_0$) is likely
to be followed by a jump back to the global minimum, which also
changes the flux by $(n+1/2)\Phi_0$. It would be useful to perform
a more careful statistical analysis of the data, which could
quantify the degree of correlation of half flux events, and
possibly shed more light on the nature of this process.

\end{widetext}

\end{document}